\documentclass[aps,prd,twocolumn,superscriptaddress]{revtex4-1}

\usepackage{longtable}
\usepackage{grffile} 
\usepackage{graphics}
\usepackage{graphicx} 
\usepackage{amsmath}    
\usepackage{hyperref}   
\usepackage{subfigure}  
\usepackage{epsfig}
\usepackage{epstopdf}
\usepackage{ulem}
\usepackage{color}

\newcommand{ \qqbar }{\mbox{$q\bar{q}$ }}

\begin{document}

\title{Effect of single string structure and multiple string interaction on strange particle production in pp collisions at $\sqrt{s}=7$ TeV}

\author{Liang Zheng}\email{zhengliang@cug.edu.cn}
\affiliation{School of Mathematics and Physics, China University of Geosciences (Wuhan),
	\break Wuhan 430074, China}
\affiliation{Key Laboratory of Quark and Lepton Physics (MOE) and \break Institute
	of Particle Physics, Central China Normal University,\break Wuhan 430079, China}

\author{Dai-Mei Zhou}\email{zhoudm@mail.ccnu.edu.cn}
\affiliation{Key Laboratory of Quark and Lepton Physics (MOE) and \break Institute
	of Particle Physics, Central China Normal University,\break Wuhan 430079, China}

\author{Zhong-Bao Yin}\email{zbyin@mail.ccnu.edu.cn}
\affiliation{Key Laboratory of Quark and Lepton Physics (MOE) and \break Institute
of Particle Physics, Central China Normal University,\break Wuhan 430079, China}

\author{Yu-Liang Yan}
\affiliation{China Institute of Atomic Energy, P. O. Box 275 (18), Beijing 102413, China}
\affiliation{Key Laboratory of Quark and Lepton Physics (MOE) and \break Institute
	of Particle Physics, Central China Normal University,\break Wuhan 430079, China}

\author{Gang Chen}
\affiliation{School of Mathematics and Physics, China University of Geosciences (Wuhan),
	\break Wuhan 430074, China}

\author{Xu Cai}
\affiliation{Key Laboratory of Quark and Lepton Physics (MOE) and \break Institute
	of Particle Physics, Central China Normal University,\break Wuhan 430079, China}

\author{Ben-Hao Sa}\email{sabh@ciae.ac.cn}
\affiliation{China Institute of Atomic Energy, P. O. Box 275 (18), Beijing 102413, China}
\affiliation{Key Laboratory of Quark and Lepton Physics (MOE) and \break Institute
	of Particle Physics, Central China Normal University,\break Wuhan 430079, China}

\date{\today}

\begin{abstract}
We present a systematic study on the strange particle production at the Large
Hadron Collider (LHC) in proton-proton (pp) collisions at $\sqrt{s}=$ 7 TeV based
on the PACIAE simulations. Two different mechanisms accounting for single
string structure variations and multiple string interactions are implemented in
the simulations. These modifications give rise to increased effective
string tension in the Lund fragmentation model and generate more strange
particles in the hadronic final state. By comparing the results with a wealth of the LHC
data, it is turned out that the inclusion of variable effective string tension
is capable to reach an improved agreement between theory and experiment,
especially on the recently observed multiplicity dependence of strangeness
enhancement in pp collisions. This approach provides us a new method to
understand the microscopic picture of the novel high multiplicity pp events
collected at the LHC in the string fragmentation framework.


\end{abstract}

\maketitle

\section{Introduction}
The production of strange hadrons plays an important role in the investigations of
properties of the strong force and the de-confined quark-gluon matter.
For the hadronic scatterings in vacuum, the strangeness content 
in the created particles is much smaller compared to the non-strange components due to the mass suppression effect in the particle production.
If the quark-gluon plasma (QGP) is produced, the chemically equilibrated
system generates a high number of strange quark pairs through the thermal
gluon fusion process and also favors the formation of multi-strange hadrons~\cite{Rafelski:1982pu,Koch:1986ud}.
 It is thus believed that the
enhancement of strange particle production, especially for multi-strange
baryons, observed in the heavy-ion collision
experiments~\cite{Andersen:1998vu,Adams:2003fy,ABELEV:2013zaa} is a characteristic
signature for the presence of QGP matter~\cite{Rafelski:2015cxa}.

Recently,striking commonalities are observed between the high multiplicity pp, proton-lead
(p-Pb) and heavy-ion collisions at LHC energies. The strangeness production, for
example, has been systematically analyzed by the ALICE collaboration with a wide
range of experimental data from pp, p-Pb to lead-lead (Pb-Pb)
collisions~\cite{ALICE:2017jyt}. A pronouncing enhancement of strange particle
relative to pion production is reported in Ref.~\cite{ALICE:2017jyt}. The
strangeness-to-pion ratio increases smoothly with the event multiplicity across
all collision systems. The magnitude of enhancement relies on the strangeness
content of the hadron. The particle ratios obtained in pp are quite similar to
those found in p-Pb at the same event multiplicities. These intriguing results
complement other
observations~\cite{Khachatryan:2010gv,CMS:2012qk,Khachatryan:2016txc,Aad:2013fja,ABELEV:2013wsa} showing similar features traditionally associated with the QGP formation and call for further theoretical investigations to understand the microscopic mechanisms that lead to these novel phenomena.

Different models are trying to interpret this universal strangeness-to-pion
ratio as a function of the event multiplicities. The statistical-thermal model
calculations are found to provide a reasonable quantitative description to the
particle ratios observed in data and suggest the suppression of strange hadron
production in pp may come from the explicit conservation of strangeness quantum
number based on the canonical approach~\cite{Vislavicius:2016rwi}. The EPOS
model~\cite{Pierog:2013ria} assumes the QGP matter is partly formed in the pp
collisions treating the interactions based on a core-corona approach. The
prediction of this model agrees qualitatively with the observed increasing trend
of the strangeness production~\cite{Aichelin:2008mi}. Another model which
qualitatively describes the data is the color ropes
mechanism~\cite{Biro:1984cf,Andersson:1991er} implemented in
DIPSY~\cite{Bierlich:2014xba} taking into account the color interactions between
strings. Other interesting extensions to the Lund string fragmentation model are
implemented in PYTHIA8~\cite{Sjostrand:2014zea} by considering thermodynamic
features of strings in a dense environment~\cite{Fischer:2016zzs} and used to
explain some of the changing flavor composition on multiplicity. As various
models can describe some key features of the data, the fundamental origin of
enhanced strangeness production in pp collisions is still largely unknown.

In this work, we introduce an effective string tension stemmed from the single
string structure variation and multiple string interaction effect to reduce the
strange quark suppression in pp collisions based on the tunneling probability in
Lund string fragmentation model. These two mechanisms of obtaining the effective
string tensions are implemented in the PACIAE Monte Carlo event
generator~\cite{Sa:2011ye} to study the strange hadron production in
$\sqrt{s}=$7 TeV pp collisions at the LHC. Systematic studies are performed to
compare the effects of different effective string tensions on the multiplicity
dependence of strange flavor composition. 

The remainder of the paper is organized as follows: in Sec.~\ref{sec:tension},
we give a short introduction on the approaches to construct the effective stirng
tension in the string fragmentation model. Detailed implementations of the
effective string tension mechanism in the PACIAE model are illustrated in
Sec.~\ref{sec:framework}. The results and comparisons to data are provided in
Sec.~\ref{sec:results}. In the end, we summarize in Sec.~\ref{sec:summary}.

\section{Effective string tension}
\label{sec:tension}
In the Lund string fragmentation model, particles are produced mainly through the
iterative breakups of the color singlet string pieces created during the
multiple parton-parton interactions in a pp collision~\cite{Andersson:1983ia}.
Given an initial string object consisting of $q_{0}$ and $\bar{q_{0}}$ endpoints
moving apart from each other, it is assumed that the original string system can
break up into two with the production of a new $q_{1}\bar{q_{1}}$ pair in the
middle of two end quarks. Therefore, one can find a new hadron formed from the
$q_{0}\bar{q_{1}}$ object, leaving $q_1$ behind which may at a later stage pair
with other iterative creations or with $\bar{q_0}$ directly. The hadron
information is then determined by the parton flavor and kinematic configurations
of its quark components. If a diquark-antidiquark pair is generated at the
breakup point instead of the \qqbar pair, a baryon can be created in the same
formalism. As required by the local flavor conservation, the new \qqbar pair is
always produced at a common vertex. Similar to the Schwinger particle production
model in electric field~\cite{Schwinger:1951nm}, virtual particles only
hadronize when the \qqbar pair tunnels out a distance $d=m_{\perp}/\kappa$. The
tunneling probability can thus be obtained as
\begin{eqnarray}
P(m_{\perp q}) & \propto & \exp(-\frac{\pi}{\kappa}m_{\perp q}^2) \nonumber \\
               & =       & \exp(-\frac{\pi}{\kappa}m_q^2) \exp(-\frac{\pi}{\kappa}p_{\perp q}^2), \label{eqn:tunnel}
\end{eqnarray}
where $\kappa$ is the string tension representing the
color force acting on the quarks in the linear confinement field. 

This formula suggests the relative production of different quark flavor depends
on the effective quark mass involved in the tunneling probability. As it is
practically hard to give the mass values theoretically, the relative production
probabilities are usually treated as empirical model parameters tuned to data.
The string tension value is often assumed to be $\kappa\approx
1\,\mathrm{GeV/fm}$~\cite{Casher:1978wy, Glendenning:1983qq} for a pure \qqbar
dipole string hadronized without interactions to its close-by neighbors.
However, for the strings created in the pp collisions at LHC energies, the
structure of a string can be much more complicated than the pure dipole state
and the hadronization of each string may not be treated independently when the
interaction between neighboring strings becomes non-negligible. To
quantitatively account for these impacts, one can apply an effective string
tension in the tunneling probability Eq.~\ref{eqn:tunnel} for the estimation of
flavor compositions.

In this work, the effects of gluon wrinkling in the string structure and the
multiple string interactions in the densely populated environment are included
in the effective string tensions, individually and taken together. The first
scheme follows the reduction of strange quark suppression
mechanism~\cite{Tai:1998hd} modeled in PACIAE~\cite{Sa:2011ye} which enhances
the string tension when radiated gluons exist on a string piece. The
parameterized effective string tension responsible for the single string structure change
can be given as:
\begin{equation}
\kappa_{eff}^{s}=\kappa_0 (1-\xi)^{-\alpha},
\label{eqn:single_kappa}
\end{equation}
where $\kappa_0$ is the pure \qqbar string tension usually set to 1 GeV/fm, $\alpha$ is a parameter
to be tuned with experimental data while $\xi$ can be parameterized as:
\begin{equation}
\xi = \frac{\ln(\frac{k_{\perp max}^2}{s_0})}{\ln(\frac{s}{s_0})+\sum_{j=gluon}\ln(\frac{k_{\perp j}^2}{s_0})},
\end{equation}
with $k_{\perp}$ being the transverse momentum of the gluons inside a dipole
string. $\sqrt{s}$ and $\sqrt{s_0}$ give the mass of the string system and a
parameter related to the typical hadron mass, respectively. This $\xi$ quantifies
the difference of a gluon wrinkled string to a pure \qqbar string. The fractal structure of a string object is dominated by the hardest gluon on the string. The quantity $\xi$ is defined to
measure the fraction of the multiplicity introduced by the hardest gluon in a string object. $(1-\xi)^{-1}$ in 
Eq.~\ref{eqn:single_kappa} thus describes the multiplicity enhancement factor of the hardest gluon to the rest of the string component and can be related to
the string tension with a scaling formula. The
value of this string tension changes on a string-by-string basis in the current
implementation and takes the string-wise fluctuations into consideration.

On the other hand, we consider the multiple string interaction effects from the
correlation of strings overlapped in a limited transverse space by
parameterizing the effective string tension in a similar spirit of the
close-packing strings discussed in Ref.~\cite{Fischer:2016zzs} as follows:
\begin{equation}
\kappa_{eff}^{m}=\kappa_0 (1+\frac{n_{MPI}-1}{1+p^{2}_{T\ ref}/p^{2}_0})^{r},
\label{eqn:global_kappa}
\end{equation}
in which $n_{MPI}$ indicates the number of multiple parton interactions in a pp
collision event and $p^{2}_{T\ ref}/p^{2}_0$ shows the transverse scale of a
typical string object relative to the proton size. The exponent $r$ is then
treated as a free parameter. Again, $\kappa_0$ provides the string tension
without any modifications and takes the same value as used in
Eq.~\ref{eqn:single_kappa}.
As larger $n_{MPI}$ leads to a denser string system in an event, $n_{MPI}$ is strongly correlated with the number of charged particles. Together with the typical string capacity factor $p_{T}^{ref}/p_{0}$ of a proton target in the transverse space, Eq.~\ref{eqn:global_kappa} then characterizes the overlap degree of string objects created in one event. Unlike the close-packing effective string tension in Ref.~\cite{Fischer:2016zzs}, $\kappa_{eff}^{m}$ is supposed to capture the overall event activity feature and provides a generic way to model the inter-string interaction effect without considering the differential picture of the string population.
This multiple string interaction triggered effective
tension applies globally to all the string objects in the same event and thus
serves as an event-wise effective string tension moderator. It is expected that
the effect of inter-string interactions can be modeled in this functional form
without considering the details of individual string dynamics.

Additionally, the string-wise single string structure change and event-wise
multiple string interaction effect can be combined together by replacing the
$\kappa_0$ in Eq.~\ref{eqn:global_kappa} with $\kappa_{eff}^{s}$. Assuming
$\kappa_0= 1$ GeV/fm, one can derive a combined effective string tension as a
product of the two:
\begin{eqnarray} 
\kappa^{s+m}_{eff} = & \kappa^{s}_{eff} (1+\frac{n_{MPI}-1}{1+p^{2}_{T\ ref}/p^{2}_0})^{r} \nonumber \\
  = & \kappa^{s}_{eff}\times\kappa^{m}_{eff}.
 \label{eqn:couple_kappa}
\end{eqnarray}

By introducing the effective string tension, the quark relative
production ratio parameters can be modified by a scaling relation as
implied in the tunneling probability accordingly.

\section{The simulation framework}
\label{sec:framework}
The variational string tension study is performed with the PACIAE simulations in
this work. The PACIAE model is a multi-purpose Monte Carlo event generator
developed to describe a wide range of collisions including hadron-hadron
interactions, hadron interactions off nuclei and nucleus-nucleus collisions. It
is built based on PYTHIA-6.4~\cite{Sjostrand:2006za} and incorporates the parton
and hadron rescattering stages to take care of the nuclear medium effects. For pp
collisions, the PACIAE model is different from PYTHIA with the capability to
carrying out parton cascade before hadronization and hadron rescattering after
hadronization. We follow the strategy in Ref.~\cite{Sa:2011ye} and
Ref.~\cite{Bierlich:2014xba} which introduce a modification on the relevant
string fragmentation parameters to account for the string tension alterations.
The following relevant string fragmentation parameters are supposed to evolve as
the string tension directly:
\begin{itemize}
\item $\rho$, strange to light quark ratio P(s)/P(u), PARJ(2) in PYTHIA; 
\item $x$, extra suppression on diquarks with strange content, PARJ(3) in PYTHIA; 
\item $y$, spin 1 to spin 0 diquark ratio, PARJ(4) in PYTHIA;
\item $\sigma$, Gaussian width of the transverse momentum distribution for primary hadrons in fragmentation, PARJ(21) in PYTHIA. 
\end{itemize}
\begin{figure}
	\includegraphics[width=0.5\textwidth]{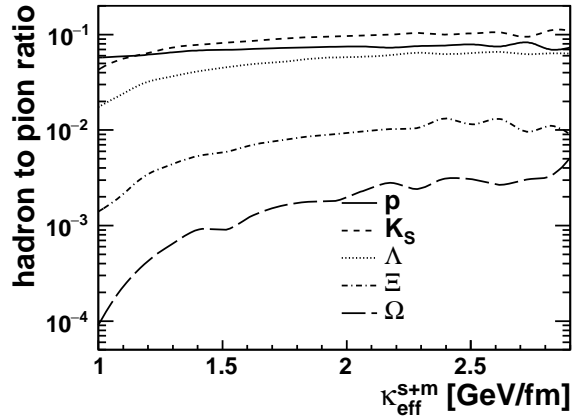}
	\caption{Relative production with respect to $\pi$ for different particles
		in the combined string tension change scheme varying with $\kappa_{eff}^{s+m}$ in $\sqrt{s}=$ 7TeV pp collisions.}
	\label{fig:particle_ratio_kappa}
\end{figure}

\begin{figure}[hbt]
	\centering
		\subfigure{
		\includegraphics[width=0.42\textwidth]{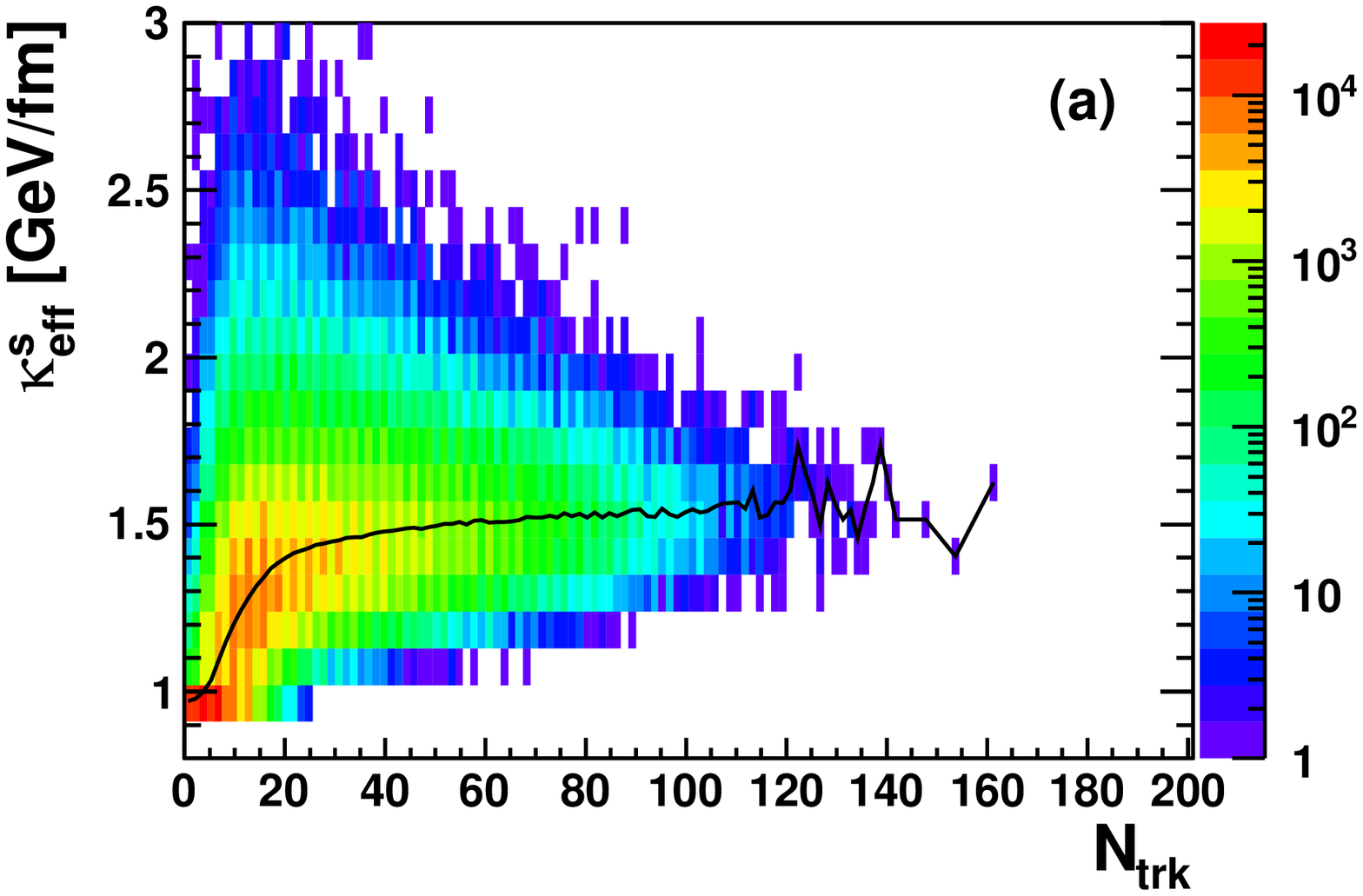}
		\label{fig:kappa_s}
	}
	\subfigure{
		\includegraphics[width=0.42\textwidth]{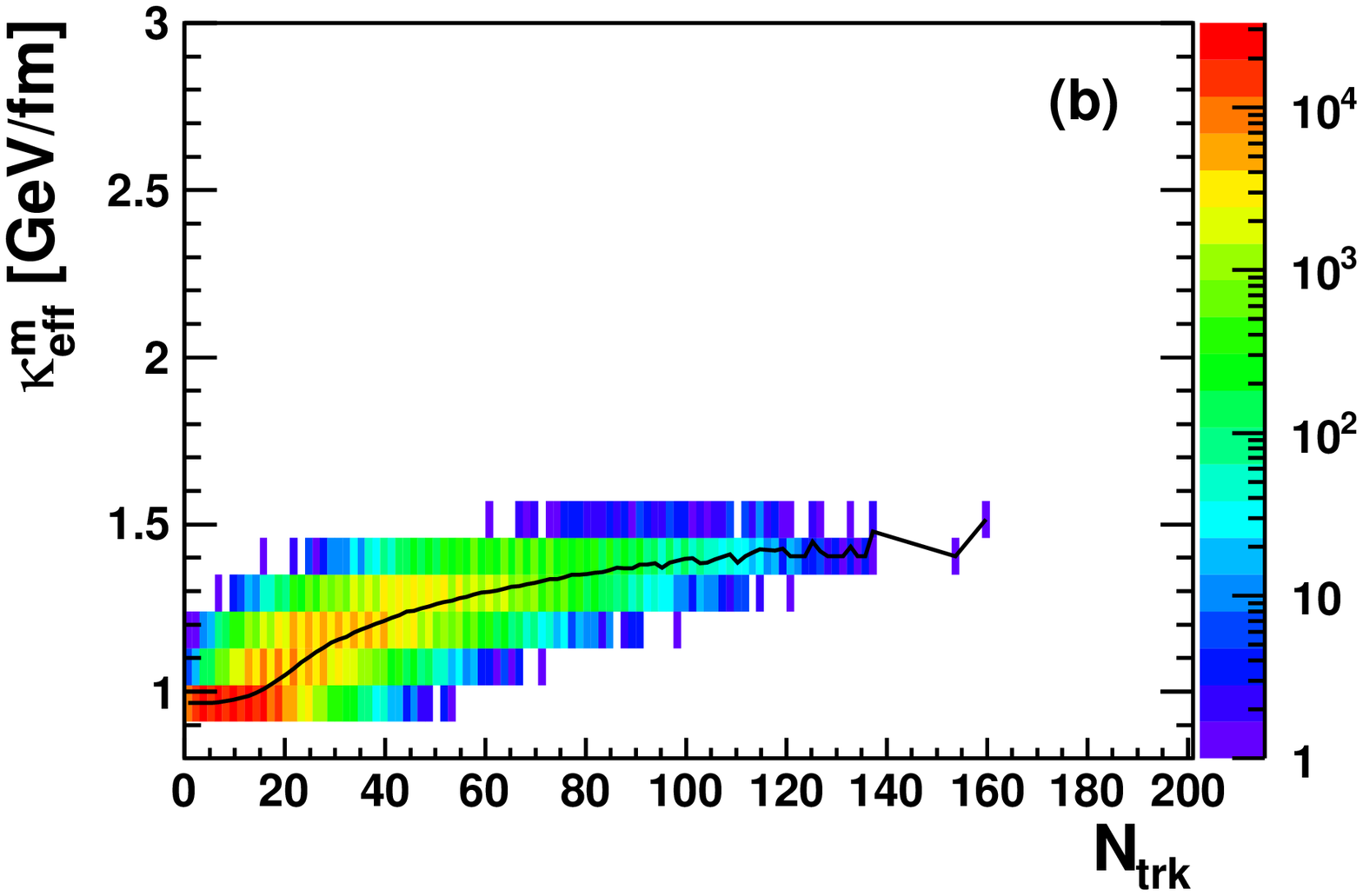}
		\label{fig:kappa_g}
	}
	\subfigure{
		\includegraphics[width=0.42\textwidth]{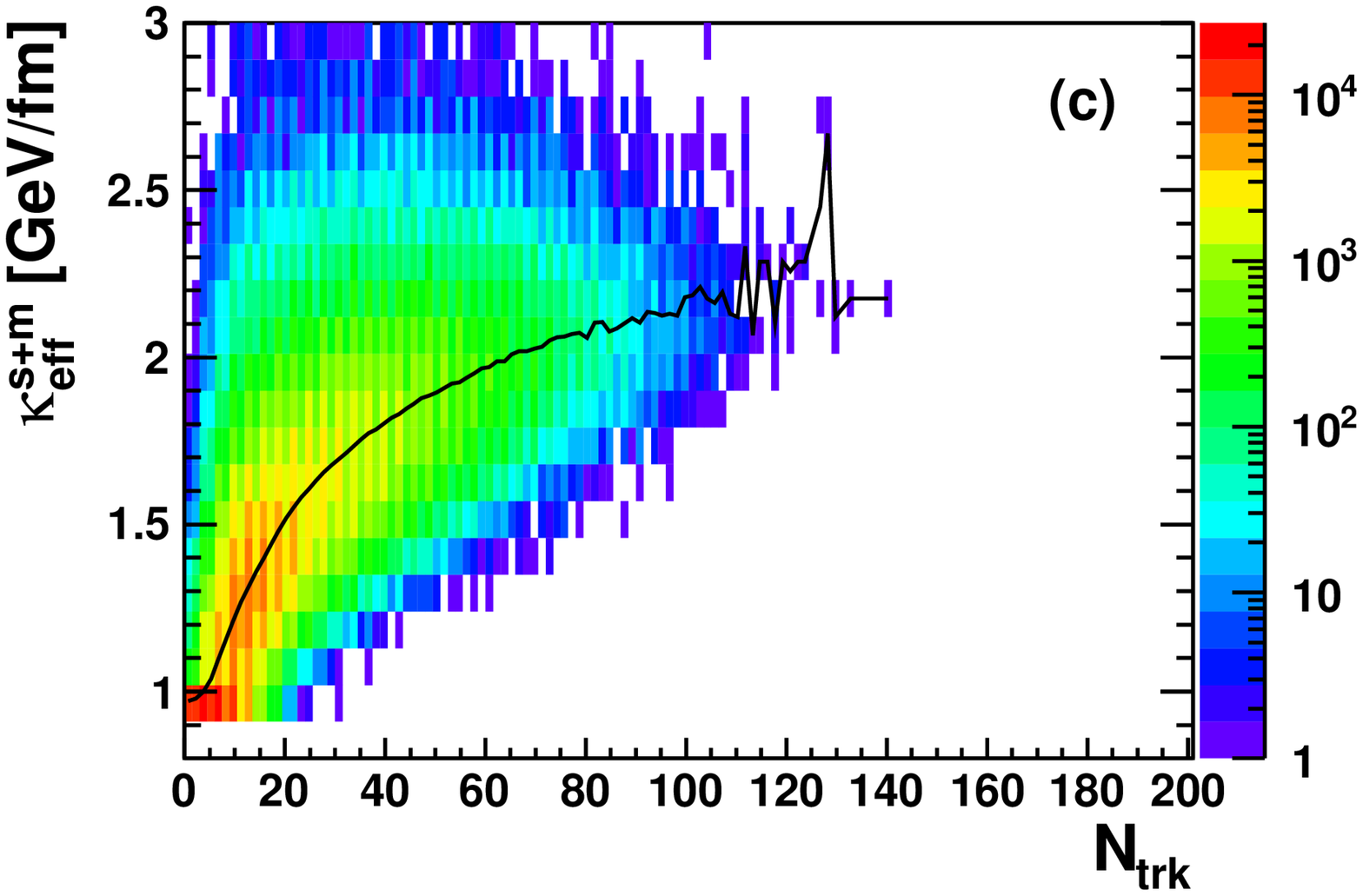}
		\label{fig:kappa_sg}
	}
	\caption{(Color online) Correlation of the effective string tension and the
	number of charged tracks accepted within the region $-3.7<\eta<-1.7$ and
$2.8<\eta<5.1$ (following the ALICE analysis convention) in pp collisions at
$\sqrt{s}=7$ TeV for single string structure change $\kappa_{eff}^{s}$ (a),
multiple string interaction effect $\kappa_{eff}^{m}$ (b) and combined string
tension $\kappa_{eff}^{s+m}$ (c), respectively. Solid line shows the average
value of the effective string tension dependent on multiplicity in each
scenario. }
	\label{fig:kappa_variation}
\end{figure}
The overall diquark to quark ratio P(qq)/P(q), $\eta$ (PARJ(1) in PYTHIA),
relies on the above parameters and changes with the string tension indirectly.
We take the pre-tuned parameters based on the inclusive measurements as the
reference values at $\kappa=\kappa_0$ and modify these parameters according to a
simple scaling indicated by Eq.~\ref{eqn:tunnel}. This means if $\kappa=\kappa_{eff}$, then
$\rho_{eff}=\rho_0^{\kappa_{0}/\kappa_{eff}}$,
$x_{eff}=x_{0}^{\kappa_{0}/\kappa_{eff}}$,
$y_{eff}=y_{0}^{\kappa_{0}/\kappa_{eff}}$,
$\sigma_{eff}=\sigma_{0}\sqrt{\kappa_{eff}/\kappa_0}$. However, the diquark to quark
ratio needs to be revised as
$\eta_{eff}=w_{eff}\beta(\frac{\eta_{0}}{w_{0}\beta})^{\kappa_{0}/\kappa_{eff}}$, 
where $w_{eff}=\frac{1+2x\rho+9y+6xy\rho+3x^2 y\rho^2}{2+\rho}$ depicts the weighting 
factor from all different kinds of diquark or quark combinations. As diquarks in the 
model are effectively generated through a stepwise popcorn mechanism~\cite{Andersson:1984af} 
from the successive production of several \qqbar pairs, $\beta$ is introduced for the 
probability to have a \qqbar fluctuation at the first place which is independent 
of the string tension.

The reference values for the above pretuned string fragmentation parameters are
taken from Perugia2011 tune~\cite{Skands:2010ak} (PARJ(1)=0.087, PARJ(2)=0.19,
PARJ(3)=0.95, PARJ(4)=0.043, PARJ(21)=0.33), which provides a reasonable
description of the inclusive measurements at the LHC energy. To describe the
charged particle density, a factor $K=0.82$ needs to be introduced in PACIAE for
the hard scattering cross sections. The fraction of diquarks from two step quark
anti-quark pair productions $\beta$ is found to be important for the
multiplicity dependence of the baryon to meson ratio. We fix $\beta=0.05$ to
give a flat $\Lambda/K_{S}^{0}$ and $p/\pi$ ratio dependent on event
multiplicity. The single string structure effective string tension
$\kappa_{eff}^{s}$ related parameters $\alpha=3.5$ and $s_{0}=0.8$
GeV are taken from earlier fit to experimental data in Ref.~\cite{Tai:1998hd}. For the multiple string interactions, we set $p^{2}_{T\ ref}/p^{2}_0=1$ and
$r=0.2$ in this work.

Figure.~\ref{fig:particle_ratio_kappa} shows different particle over $\pi$
ratios varying with the string tension based on the parameter setup discussed
above for pp collisions at $\sqrt{s}=$ 7 TeV. The result is given by the
combined effective string tension $\kappa_{eff}^{s+m}$ simulation with both
string structure change and multiple string interactions shown in
Eq.~\ref{eqn:couple_kappa}. We find a rapid increase with $\kappa_{eff}^{s+m}$
in strange particle relative production especially for the multi-strange
particles. Due to the choice of a small $\beta$ value, $p/\pi$ ratio barely
changes with respect to the string tension.

We also explore the string tension varying with the event multiplicity in
different scenarios as shown in Fig.~\ref{fig:kappa_variation}. The event-wise
track numbers at forward rapidities are used as the event activity estimator. In
this comparison, one can observe that there is a correlation between the
effective string tension and the event multiplicity. For the single string
structure related effective string tension $\kappa_{eff}^{s}$, a rapid increase
of string tension is expected in the low multiplicity region due to stronger
gluon radiations. The maximum gluon radiation energy is constrained by the
allowed phase space. For the events with high multiplicities, the effects from
the single string structure change may not be important any more. Accordingly,
if we switch to the multiple string interaction case, the average string tension $\kappa_{eff}^{m}$
of an event is rising as the multiplicity with rather small variations. By
combining the effects of the two modifications in the effective string tension $\kappa_{eff}^{s+m}$,
we get an even larger average tension as the coupling of two scenarios will
amplify this effect.

\section{Results}
\label{sec:results}
\begin{figure*}[hbt!]
	\centering
	\subfigure{
		\includegraphics[width=0.45\textwidth]{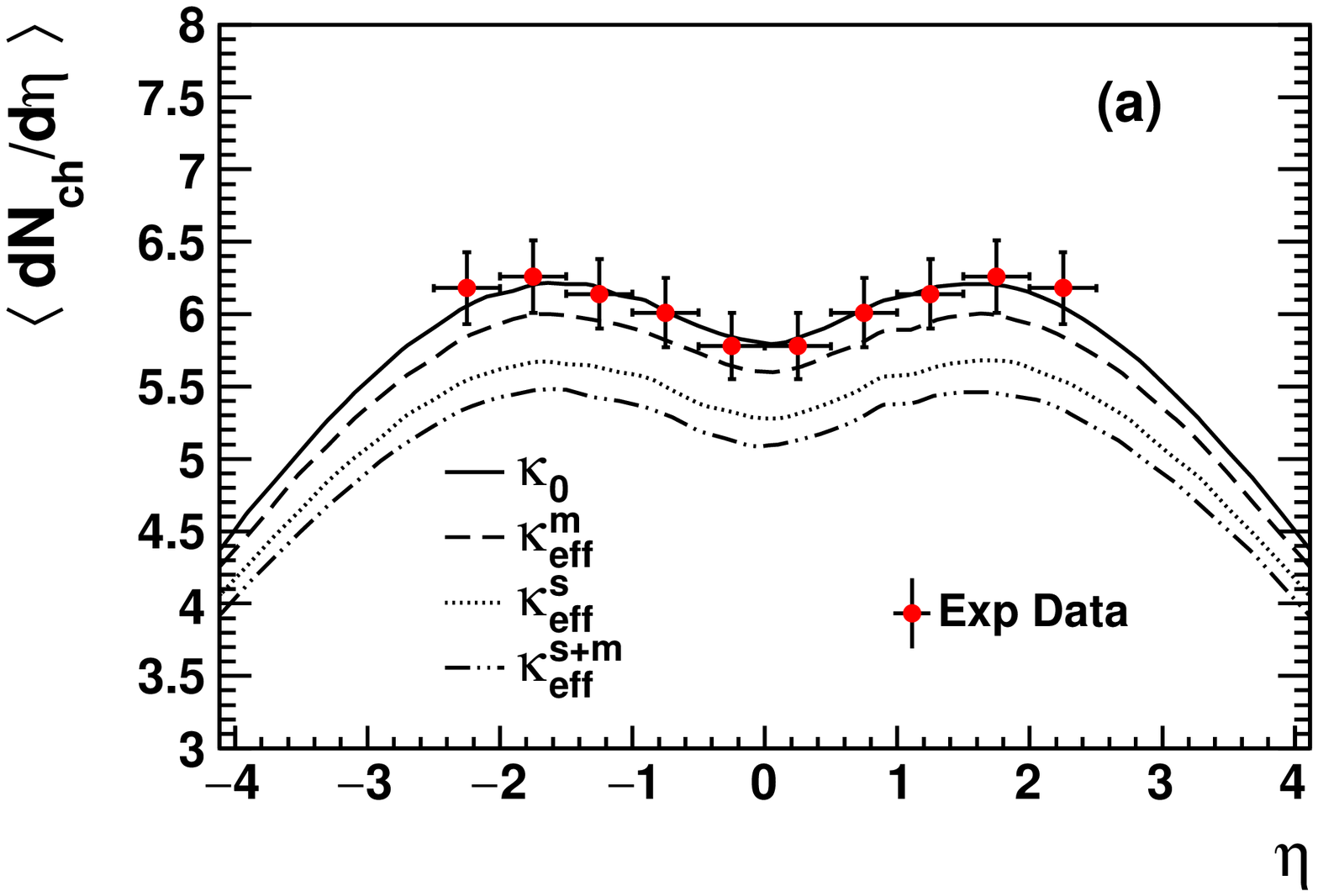}
		\label{fig:dNdEta}
	}
	\subfigure{
		\includegraphics[width=0.45\textwidth]{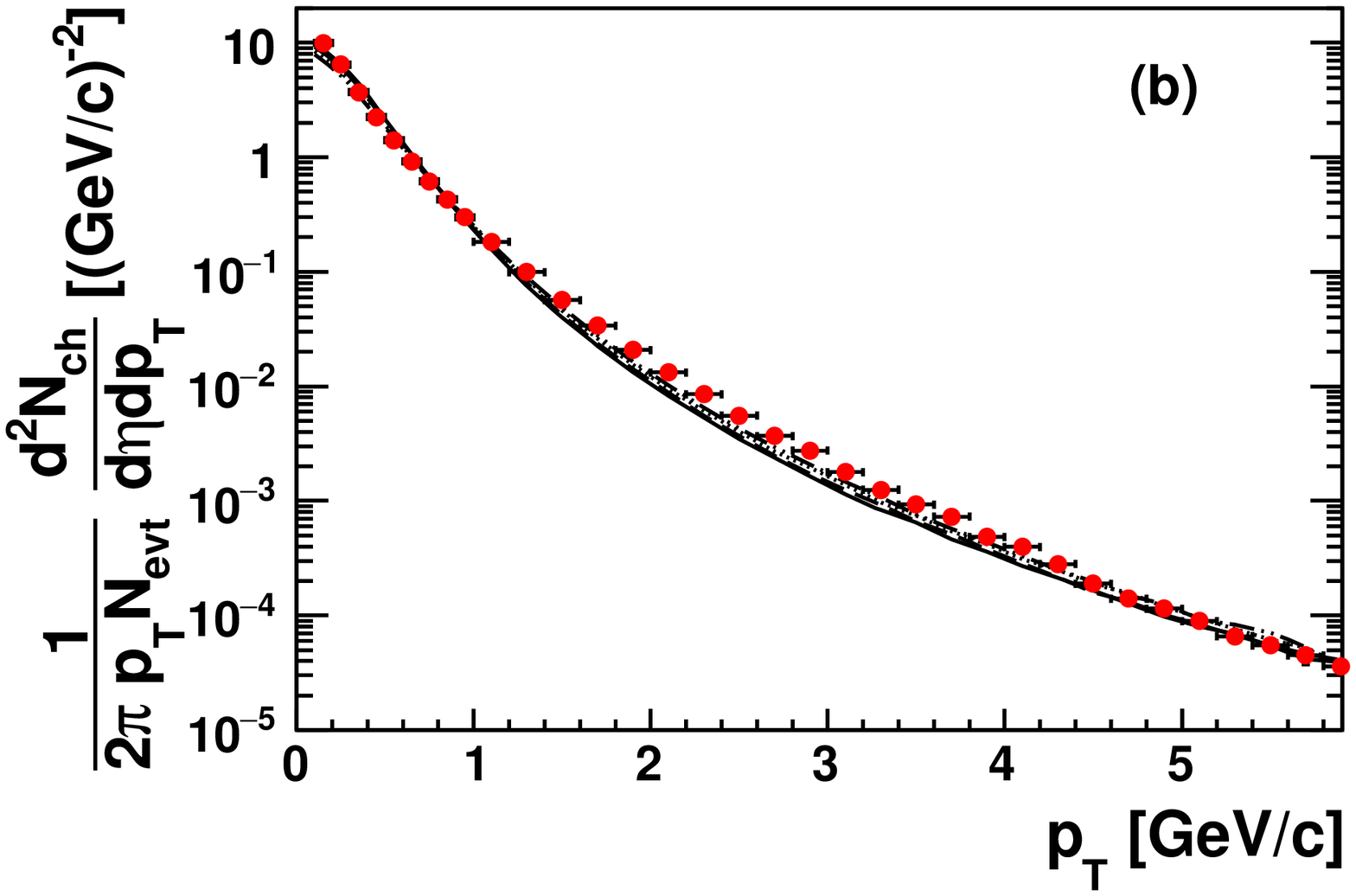}
		\label{fig:dNdPt}
	}
	\subfigure{
		\includegraphics[width=0.45\textwidth]{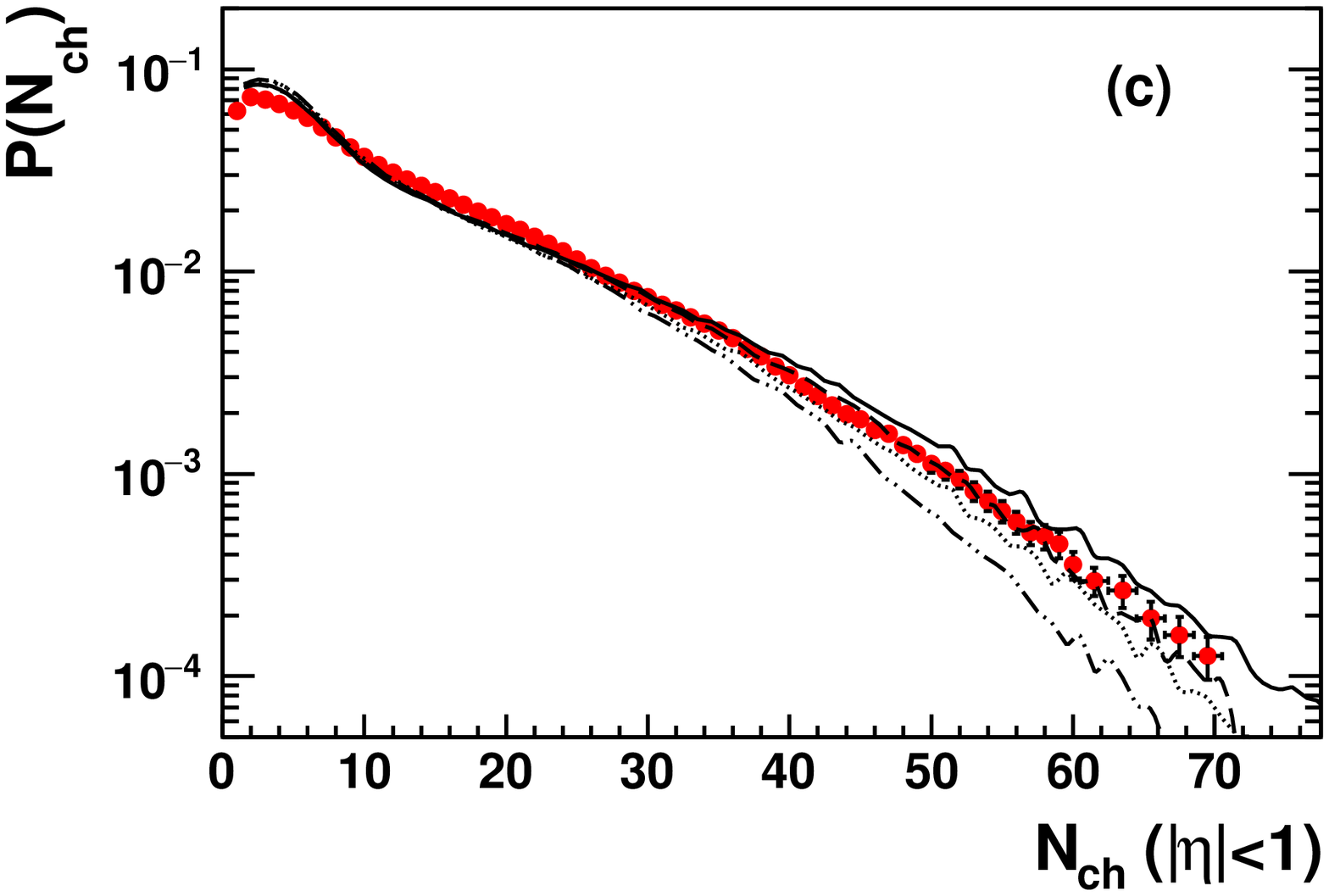}
		\label{fig:mult}
	}
	\caption{(Color online) Charged particle distributions shown in pseudorapidity (a), transverse momentum (b) and multiplicity (c) spaces for pp collisions at $\sqrt{s}=7$ TeV. Mid-rapidity charged particle numbers are obtained for $|y|<1$. Four different scenarios are shown for the $\kappa_{0}$, $\kappa_{eff}^{s}$, $\kappa_{eff}^{m}$ and $\kappa_{eff}^{s+m}$ string tension setup. The experimental data are taken from~\cite{Khachatryan:2010us,Aamodt:2010pp}. }
	\label{fig:charged_particle}
\end{figure*}

To investigate the effects of the different effective string tension scenarios, we focus on
the pp collisions with $\sqrt{s}=7$ TeV. The inclusive measurements on the
charged particle pseudo-rapidity, transverse momentum and multiplicity
distributions are presented in Fig.~\ref{fig:charged_particle}. Model predictions
with default string tension $\kappa_{0}$,
single string structure change $\kappa_{eff}^{s}$ and multiple string
interactions $\kappa_{eff}^{m}$ are shown in the solid, dashed and dotted lines,
respectively. The combination of two varying string tension scenarios
$\kappa_{eff}^{s+m}$ is shown with the dash-dotted line. It is shown in this
comparison the charged density decreases as the effective string tension
increases. This is not beyond expectations in the model as increasing string
tension makes a particle more difficult to tunnel out. This effect can also be
identified in the mid-rapidity event multiplicity distribution that high
multiplicity events become rare in $\kappa_{eff}^{s}$ and $\kappa_{eff}^{s+m}$
simulations shown in Fig.~\ref{fig:mult}. However, the impact on the charged particle transverse momentum
distribution is hardly visible with the inclusion of different string tension
parameterizations indicated by Fig.~\ref{fig:dNdPt}.

The multiplicity dependent studies are made with the event classifier counting
charged particle number accepted in the pseudo-rapidity region $-3.7<\eta<-1.7$
and $2.8<\eta<5.1$ following the ALICE data analysis method. Based on the
multiplicity distribution of the charged particle in forward rapidity region, the same event sample is divided into ten event classes with the same percentile
definition as in Ref.~\cite{ALICE:2017jyt}. For each
event class, one can estimate the charged particle pseudopraidity density within
$|\eta|<0.5$ to represent the event activity and study the per-rapidity strange
particle production in $|y|<0.5$. The average charged densities in each event
class for different string tension implementations are presented in
Tab.~\ref{tab:EvtClas}. It can be seen that the average charged densities in each
event class are close to the measured data for variations of effective string
tensions. In event class I, the
average charged density with large effective string tensions $\kappa_{eff}^{s+m}$ becomes about 20\% smaller than the case with
$\kappa_{0}$ string tension assumption. This is also consistent with our observation
on the reduction of high multiplicity events as shown in Fig.~\ref{fig:mult} with
increasing string tension.

\begin{longtable*}[H]{l|c c c c c c c c c c  }
	\caption{ Charge density in each event class $<dN_{ch}/d\eta>_{|\eta|<0.5}$. Data are taken from Ref.~\cite{ALICE:2017jyt}. \label{tab:EvtClas}} \\ \hline \hline
	Event Class & I & II & III & IV & V & VI & VII & VIII & IX & X \\ \hline
	$\sigma/\sigma_{INEL}$  & 0-0.95\% & 0.95-4.7\% & 4.7-9.5\% & 9.5-14\% & 14-19\% & 19-28\% & 28-38\% & 38-48\% & 48-68\% & 68-100\% \\
	Exp Data & $21.3\pm 0.6$ &  $16.5\pm 0.5$  & $13.5\pm 0.4$ & $11.5\pm 0.3$ & $10.1\pm 0.3$ & $8.45\pm 0.25$ & $6.72\pm 0.21$ & $5.40\pm 0.17$ & $3.90\pm 0.14$ & $2.26\pm 0.12$ \\ 
	$dN_{ch}/d\eta$($\kappa_0$) & 24.1 & 18.5 & 14.8 & 12.4 & 10.5 & 8.4 & 6.4 & 4.8 & 3.4 & 1.9 \\
	$dN_{ch}/d\eta$($\kappa_{eff}^{m}$) & 22.1 & 17.2 & 14.0 & 11.8 & 10.1 & 8.1 & 6.1 & 4.7 & 3.4 & 1.9 \\
	$dN_{ch}/d\eta$($\kappa_{eff}^{s}$) & 21.5 & 16.5 & 13.1 & 11.0 & 9.5 & 7.7 & 5.8 & 4.3 & 3.1 & 1.8 \\
	$dN_{ch}/d\eta$($\kappa_{eff}^{s+m}$) & 19.8 & 15.4 & 12.4 & 10.4 & 9.1 & 7.4 & 5.6 & 4.3 & 3.1 & 1.8 \\
	\hline \hline

\end{longtable*}
\begin{figure*}[hbt!]
	\centering
	\subfigure{
		\includegraphics[width=0.45\textwidth]{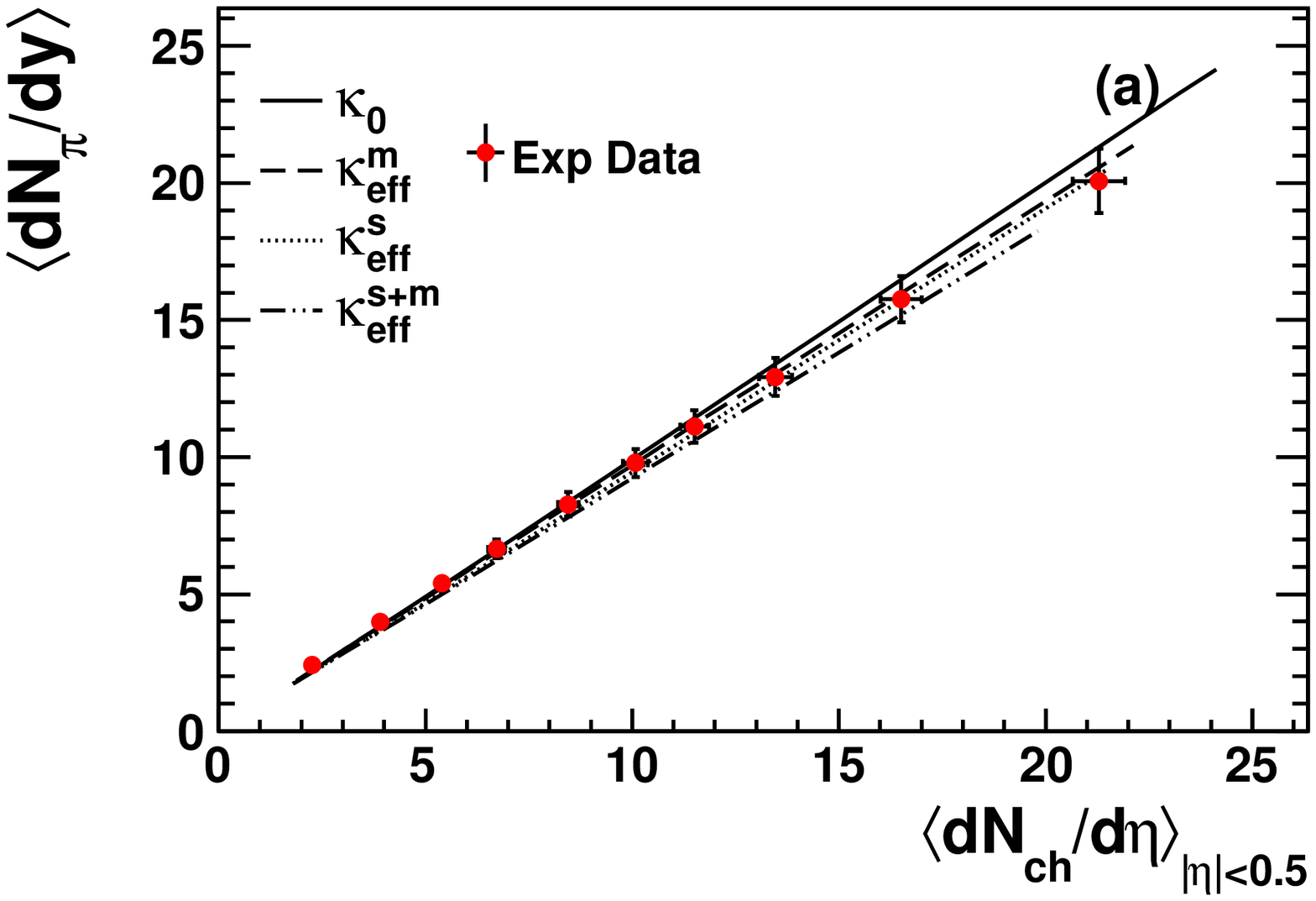}
		\label{fig:Pi_integral_mult}
	}
	\subfigure{
		\includegraphics[width=0.45\textwidth]{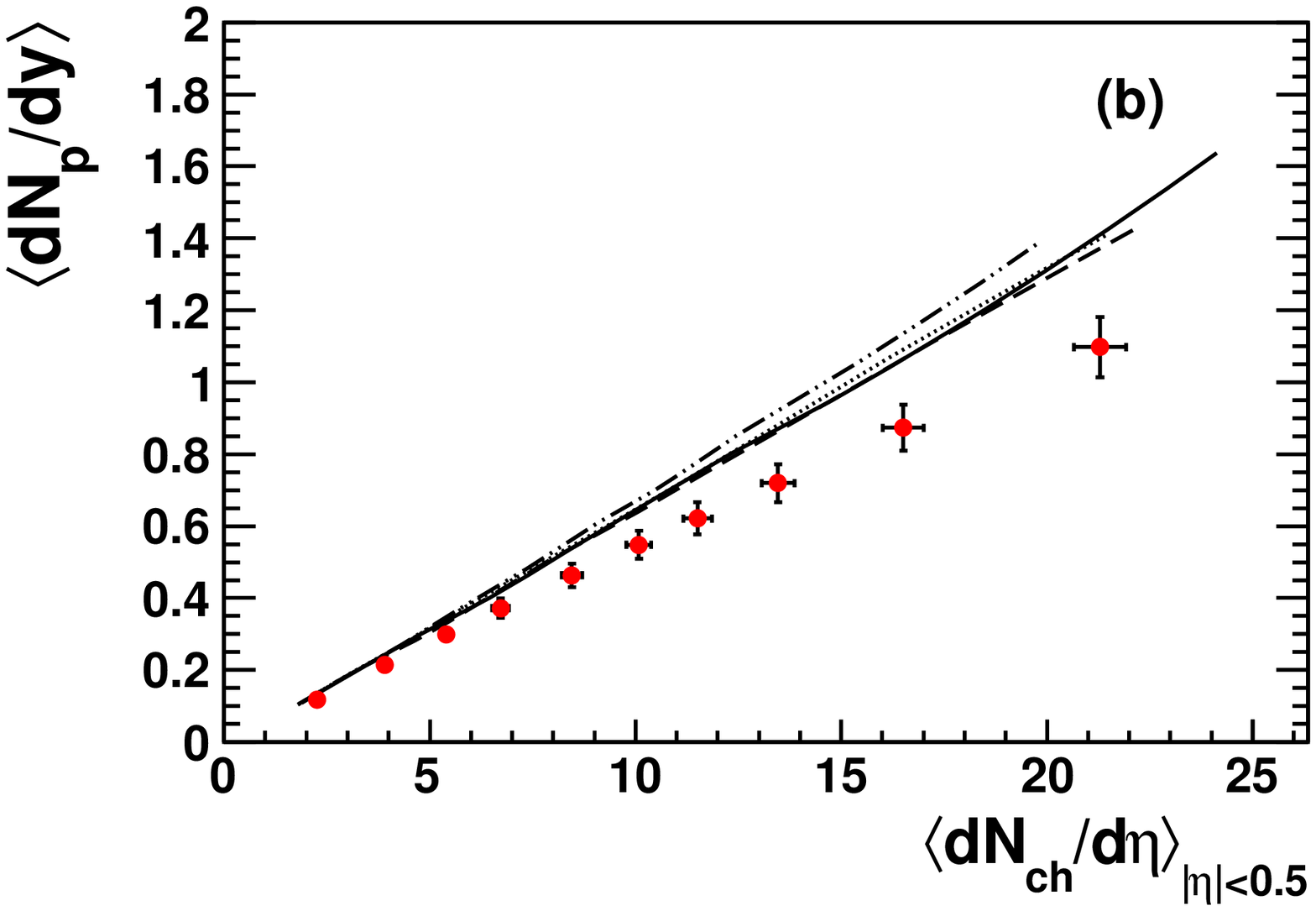}
		\label{fig:P_integral_mult}
	}
	\caption{(Color online) Pion (a) and proton (b) integrated yield varying with event multiplicity in pp collisions at $\sqrt{s}=7$ TeV for different scenarios showing: constant string tension setup ($\kappa_0$, solid line), single string-wise tension setup ($\kappa_{eff}^{s}$, dotted line), multiple string interaction tension setup ($\kappa_{eff}^{m}$, dashed line) and combined string tension setup ($\kappa_{eff}^{s+m}$, dash-dotted line). The experimental data are taken from~\cite{ALICE:2017jyt}. }
	\label{fig:pion_proton_integral_mult}
\end{figure*}
We perform an examination on the multiplicity dependence of
charged pion and proton yields as shown in
Fig.~\ref{fig:pion_proton_integral_mult}. The results suggest that the total
pion production in each event class only slightly decreases with the increasing
string tension. It is then not surprising to see the  $\pi$ yield from all four
scenarios are in reasonable agreement with the experimental data. On the other
hand, the integrated proton yield from model is higher than data even with the
constant string tension $\kappa_{0}$. All effective string tension variation
scenarios introduce a negligible increase to the integrated proton yield
comparing with the constant string tension result. It is due to our setup of the
small $\beta$ parameter which determines how strong the diquark to quark
production ratio relies on the string tension.
\begin{figure*}[hbt!]
	\centering
	\subfigure{
		\includegraphics[width=0.45\textwidth]{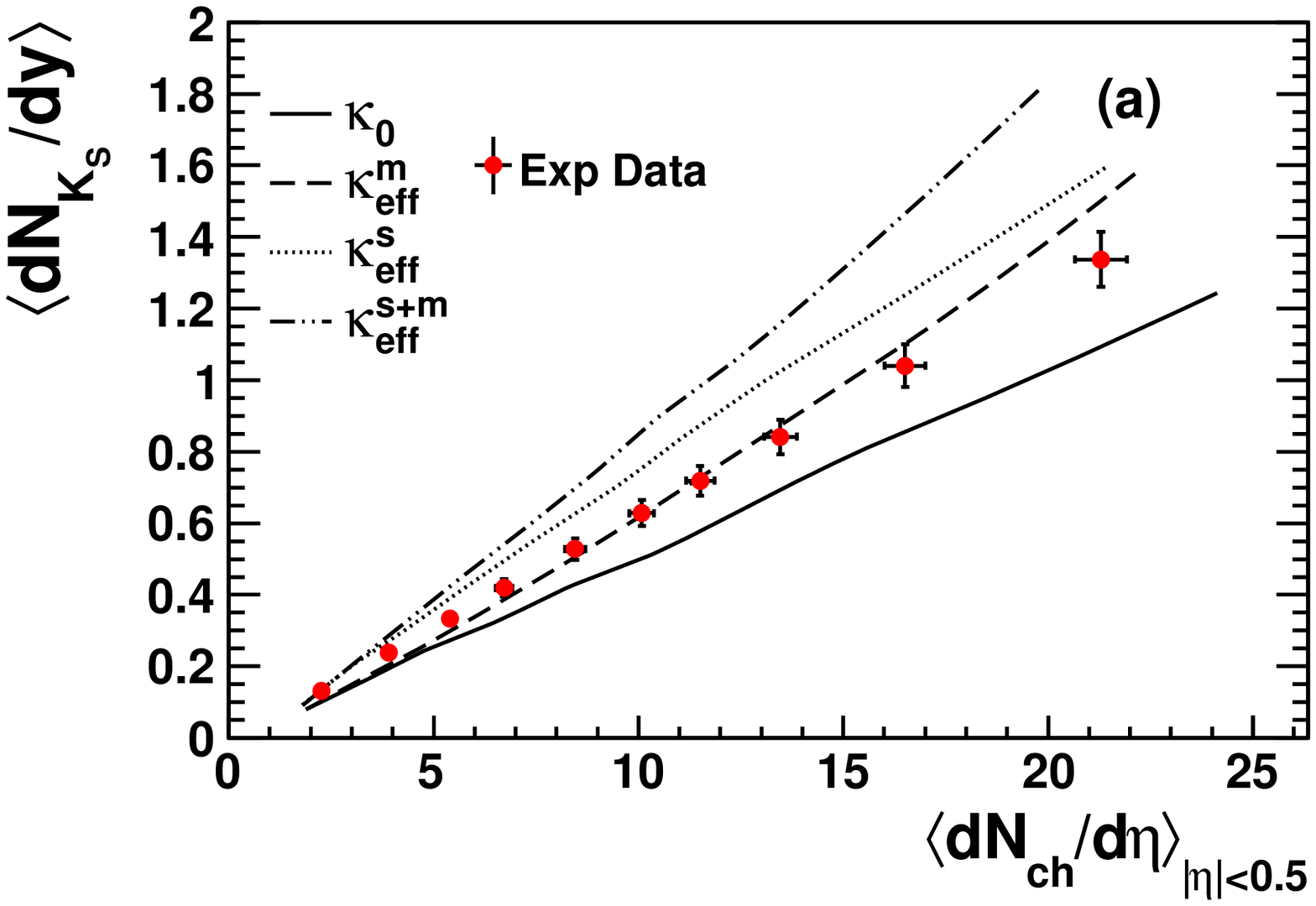}
		\label{fig:KS_integral_mult}
	}
	\subfigure{
		\includegraphics[width=0.45\textwidth]{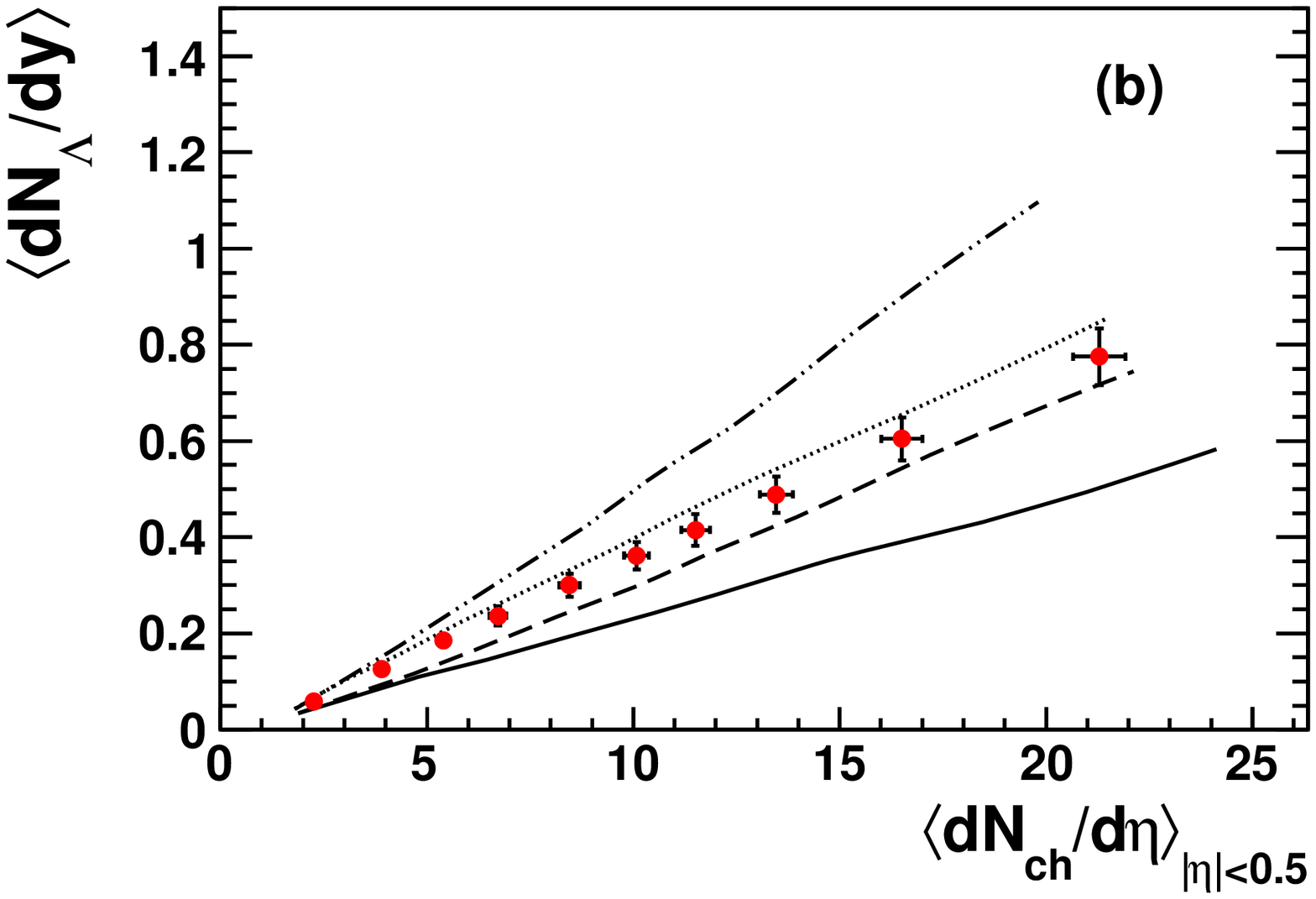}
		\label{fig:L_integral_mult}
	}
	\subfigure{
		\includegraphics[width=0.45\textwidth]{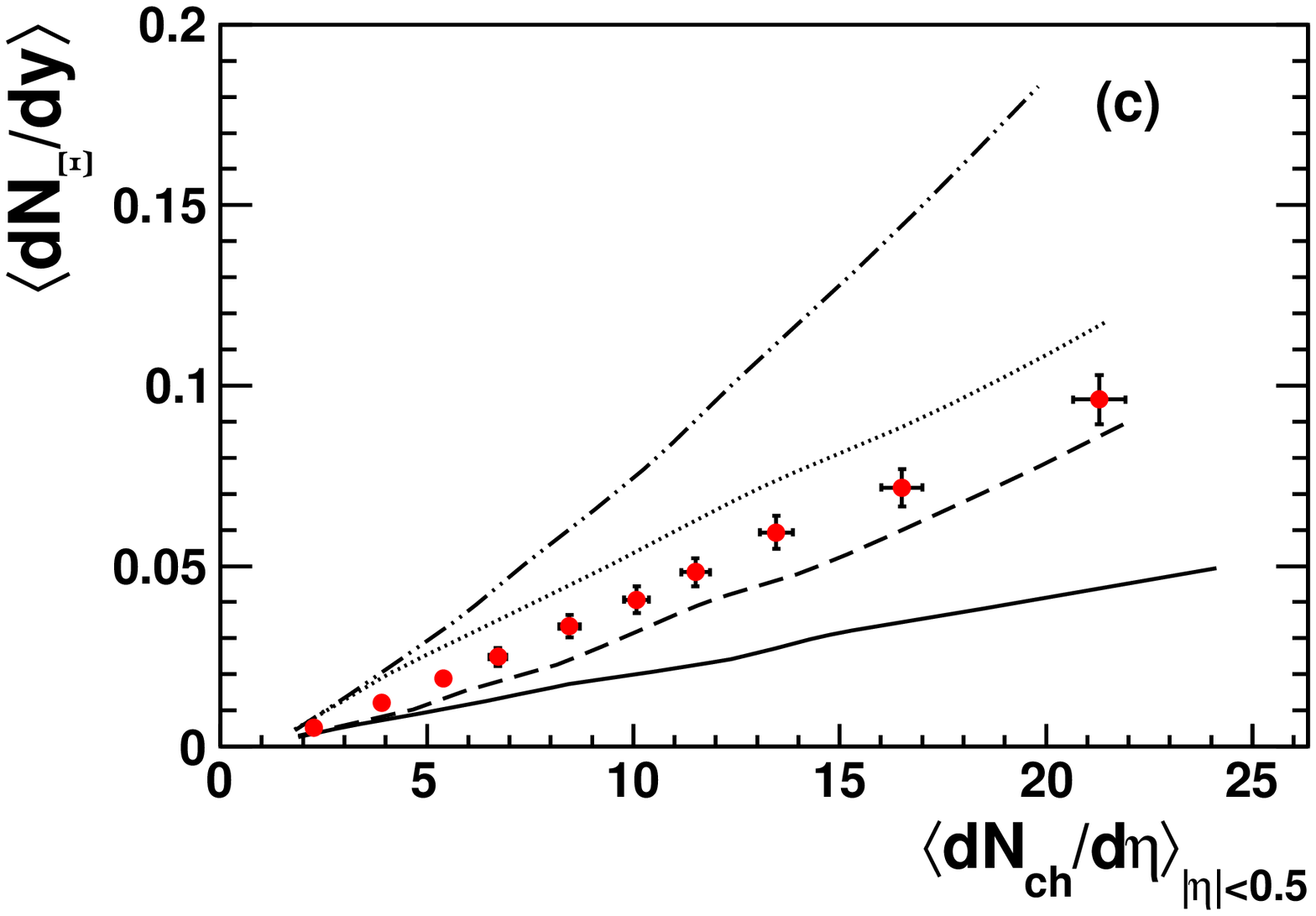}
		\label{fig:Xi_integral_mult}
	}
	\subfigure{
		\includegraphics[width=0.45\textwidth]{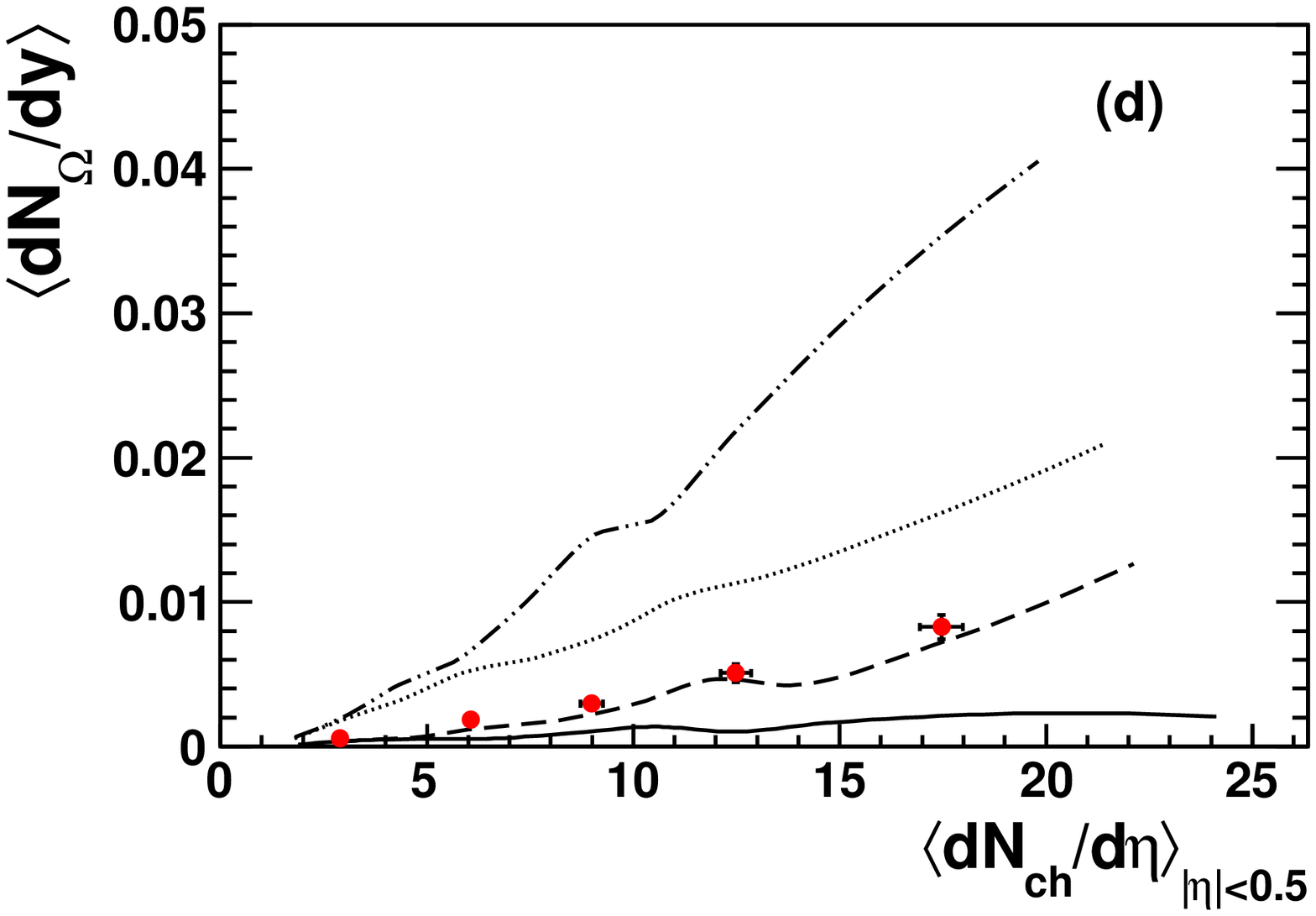}
		\label{fig:Omega_integral_mult}
	}
	\caption{(Color online) Strange particle integrated yield varying with event multiplicity in pp collisions at $\sqrt{s}=7$ TeV for $K_S^{0}$ (a), $\Lambda$ (b), $\Xi$ (c) and $\Omega$ (d).  Four different scenarios are shown with constant string tension setup ($\kappa_0$, solid line), single string-wise tension setup ($\kappa_{eff}^{s}$, dotted line), multiple string interaction tension setup ($\kappa_{eff}^{m}$, dashed line) and combined string tension setup ($\kappa_{eff}^{s+m}$, dash-dotted line). The experimental data are taken from~\cite{ALICE:2017jyt}. }
	\label{fig:strange_integral_mult}
\end{figure*}

\begin{figure*}[hbt!]
	\centering
	\subfigure{
		\includegraphics[width=0.45\textwidth]{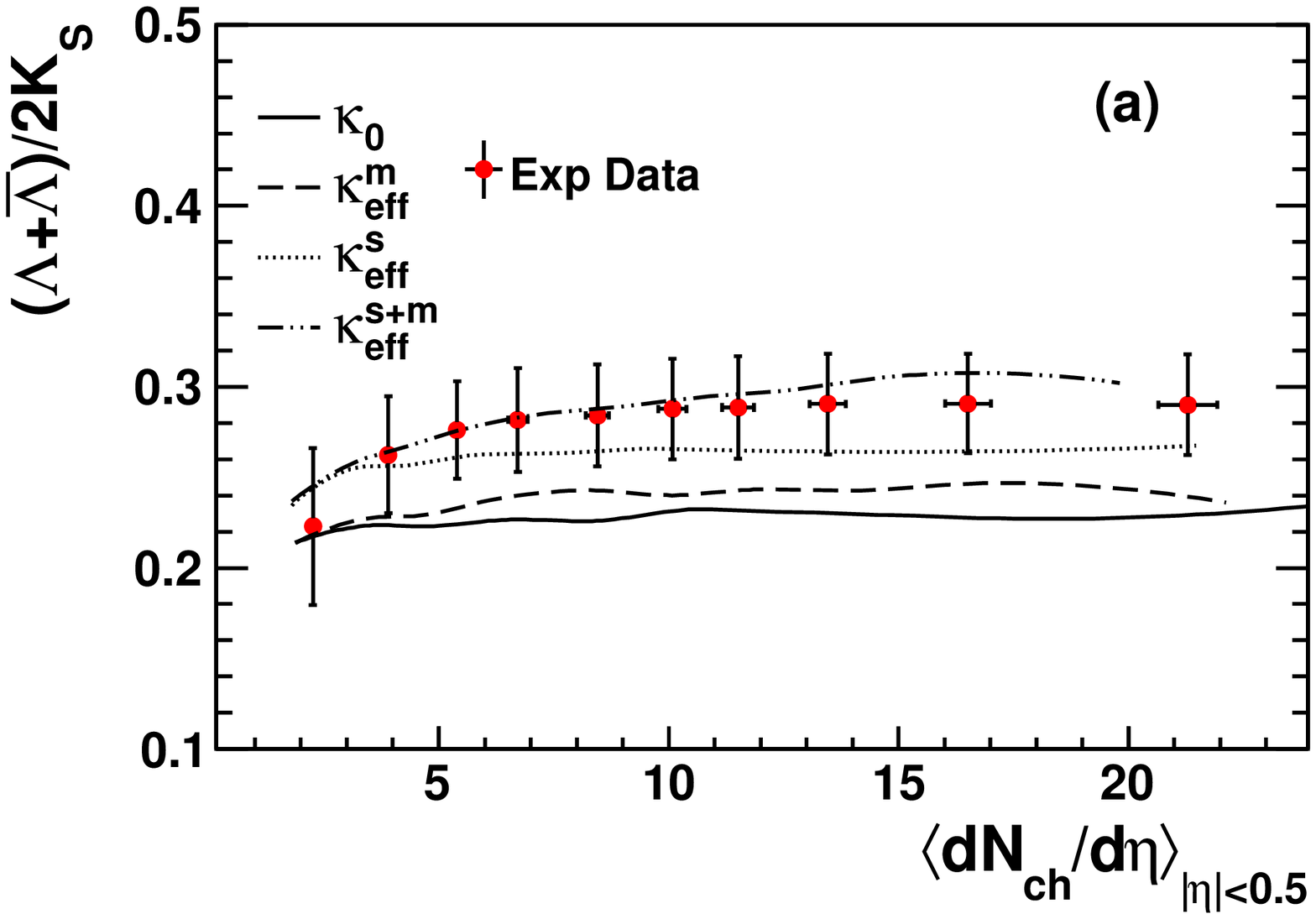}
		\label{fig:LOverK_mult}
	}
	\subfigure{
		\includegraphics[width=0.45\textwidth]{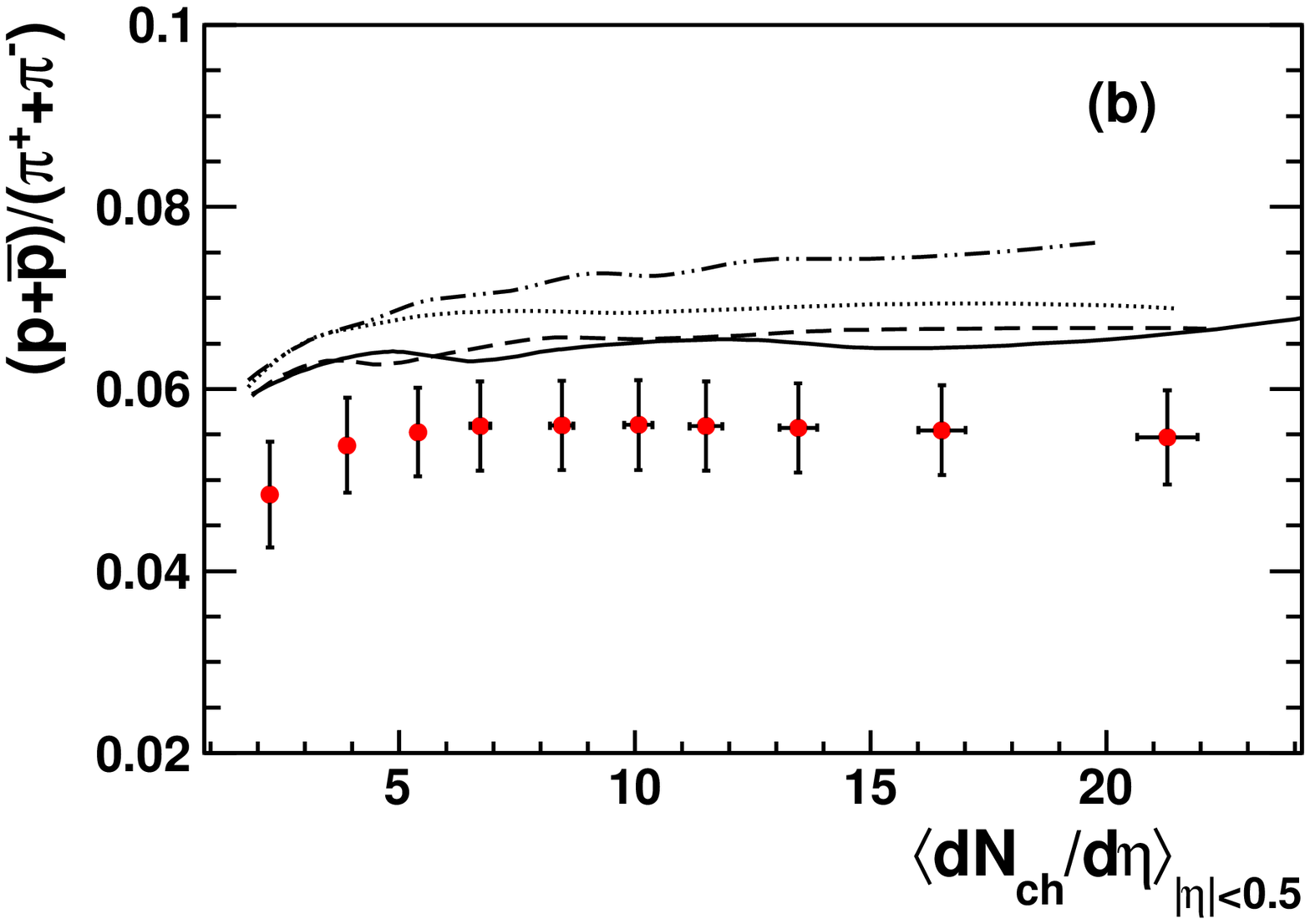}
		\label{fig:POverPi_mult}
	}
	\caption{(Color online) Baryon over meson ratio varying with multiplicity in pp collisions at $\sqrt{s}=7$ TeV. Four different scenarios are shown with constant string tension setup ($\kappa_0$, solid line), single string-wise tension setup ($\kappa_{eff}^{s}$, dotted line), multiple string interaction tension setup ($\kappa_{eff}^{m}$, dashed line) and combined string tension setup ($\kappa_{eff}^{s+m}$, dash-dotted line). The experimental data are taken from~\cite{ALICE:2017jyt}. }
	\label{fig:BOverM_mult}
\end{figure*}
\begin{figure*}[hbt!]
	\centering
	\subfigure{
		\includegraphics[width=0.45\textwidth]{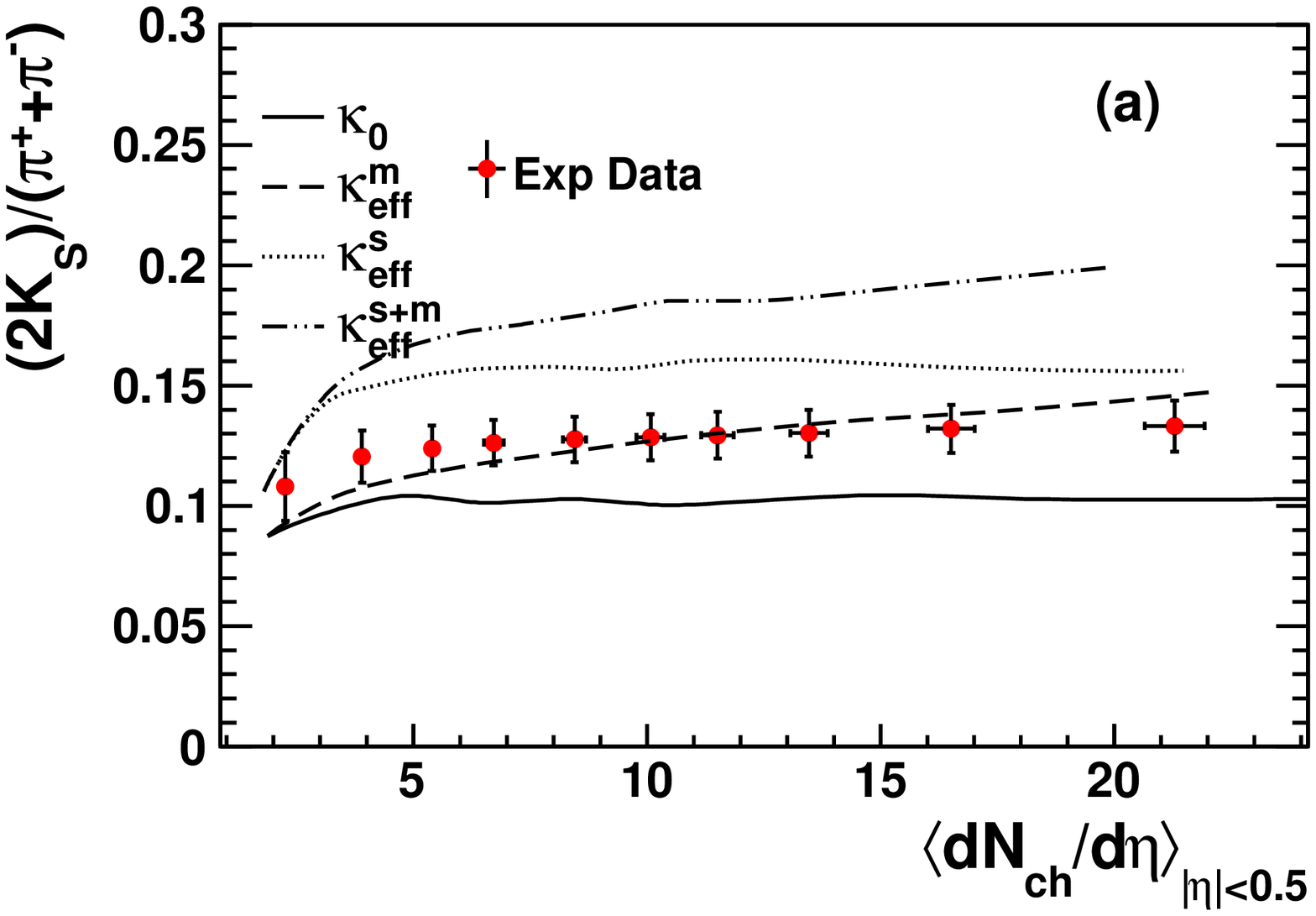}
		\label{fig:KSOverPi_mult}
	}
	\subfigure{
		\includegraphics[width=0.45\textwidth]{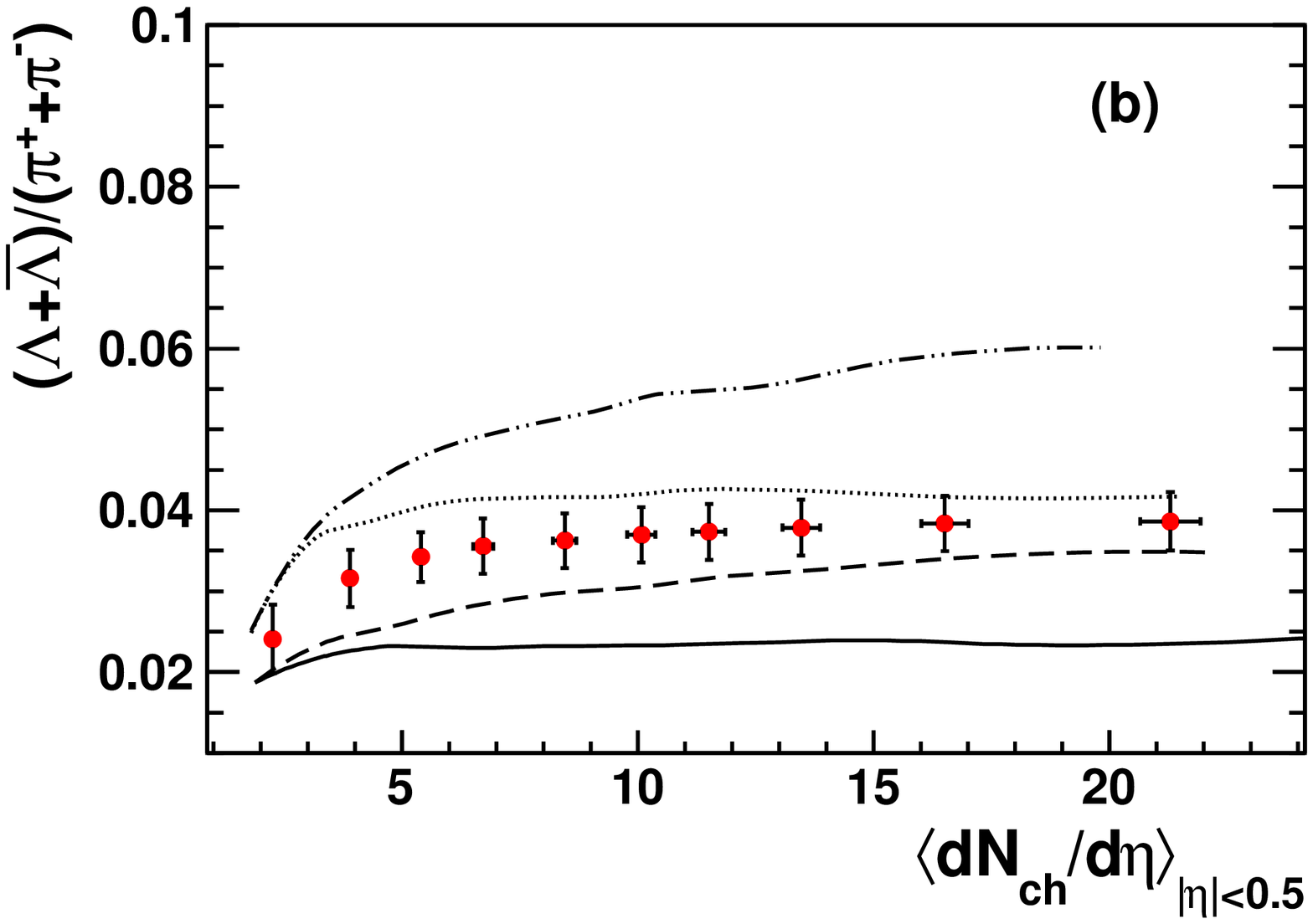}
		\label{fig:LOverPi_mult}
	}
	\subfigure{
		\includegraphics[width=0.45\textwidth]{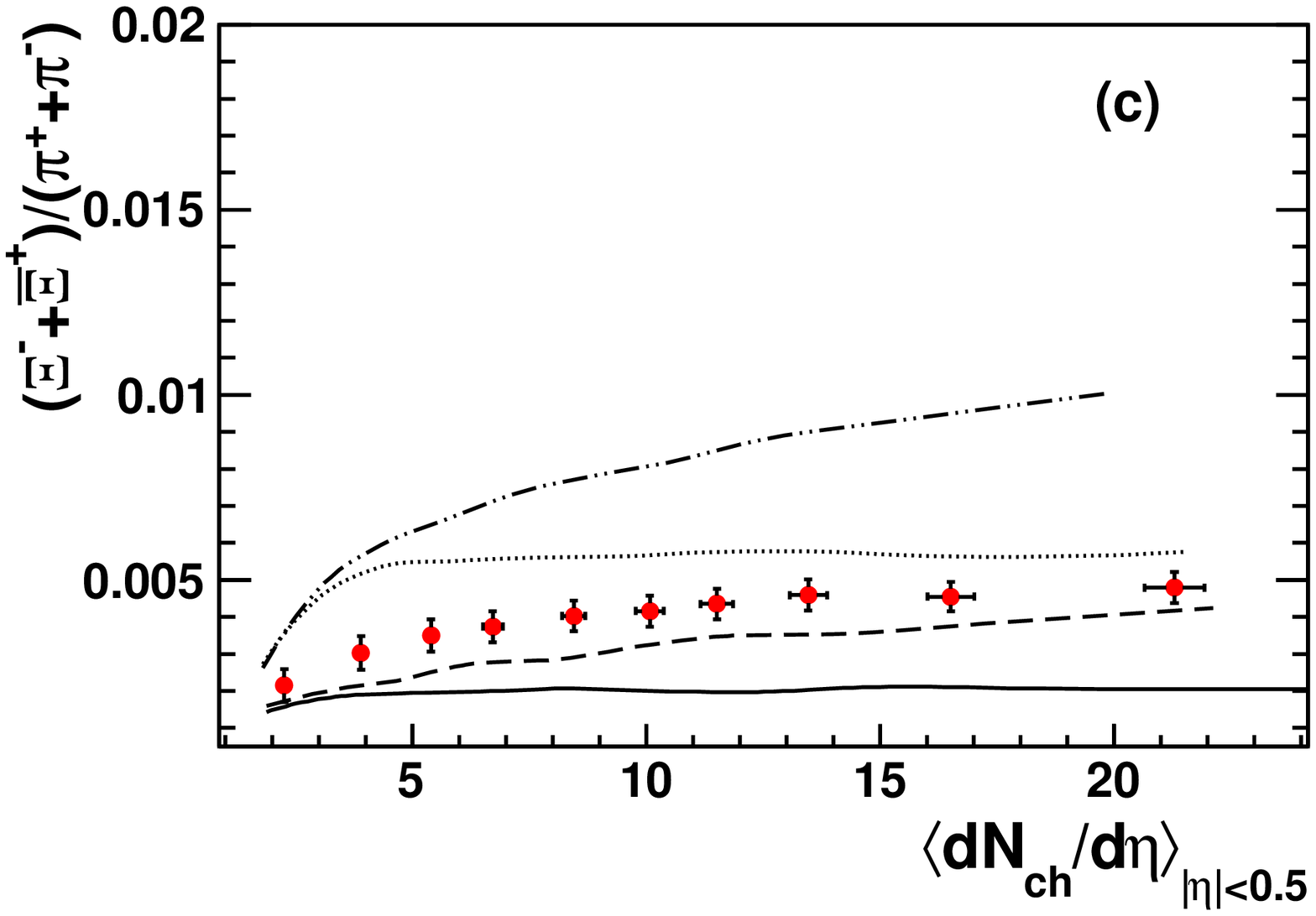}
		\label{fig:XiOverPi_mult}
	}
	\subfigure{
		\includegraphics[width=0.45\textwidth]{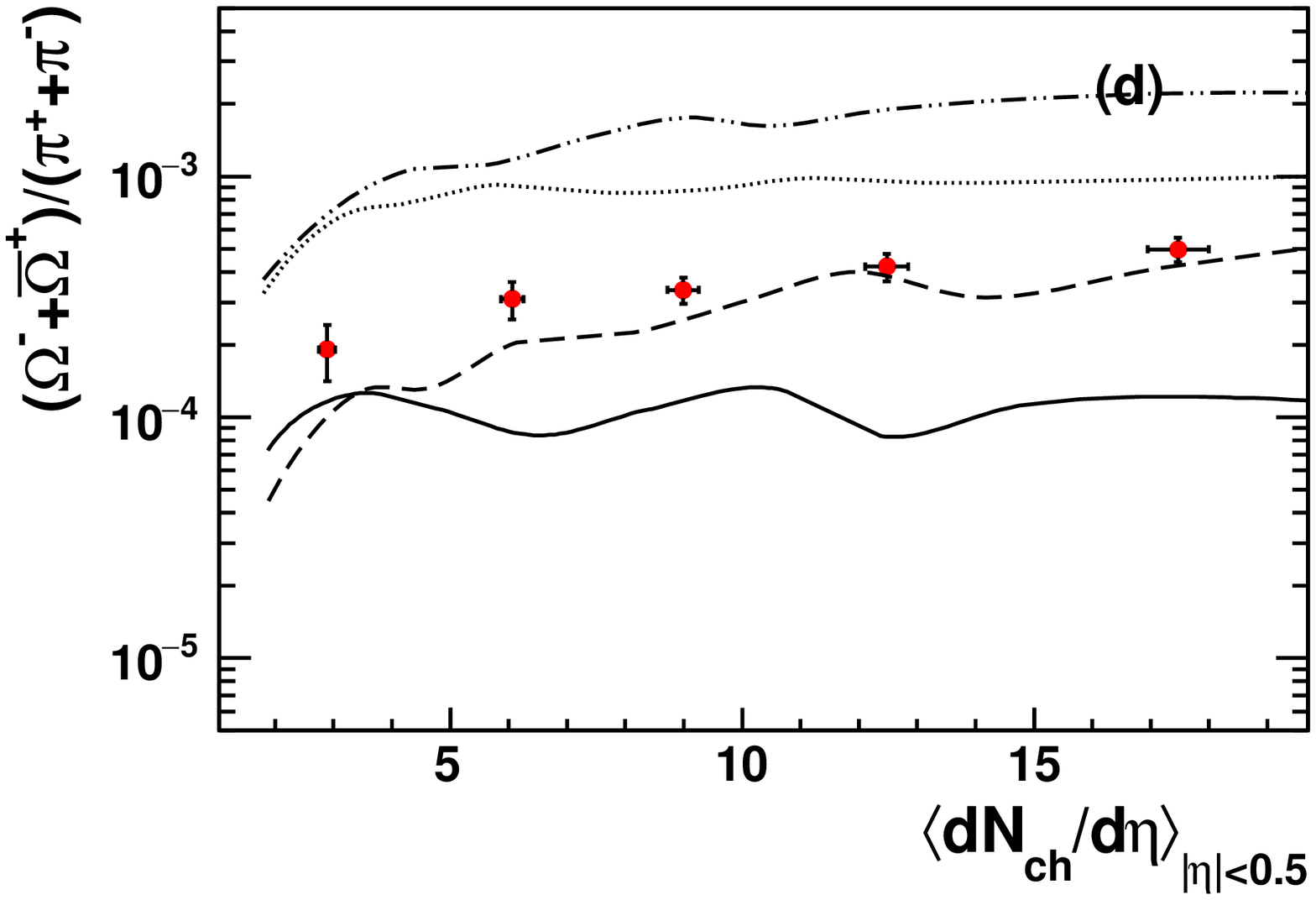}
		\label{fig:OmegaOverPi_mult}
	}
	\caption{(Color online) Strange particle over pion ratio varying with event multiplicity in pp collisions at $\sqrt{s}=7$ TeV.  Four different scenarios are shown with constant string tension setup ($\kappa_0$, solid line), single string-wise tension setup ($\kappa_{eff}^{s}$, dotted line), multiple string interaction tension setup ($\kappa_{eff}^{m}$, dashed line) and combined string tension setup ($\kappa_{eff}^{s+m}$, dash-dotted line). The experimental data are taken from~\cite{ALICE:2017jyt}. }
	\label{fig:strangeOverPion_mult}
\end{figure*}

In Fig.~\ref{fig:strange_integral_mult}, we show a comparison of simulation
results on multiplicity dependence of strangeness production with experimental
data. As can be seen in Fig.~\ref{fig:KS_integral_mult}, $K_S^{0}$ production is
enhanced with the inclusion of a stronger string tension. Comparing the
$K_S^{0}$ production in the events with largest charged particle density between
the $\kappa_{0}$ and $\kappa_{eff}^{s+m}$ scenario, the yield is increased by a
factor of 50\%. The impact of the string tension change on strange baryon
productions is much more pronounced than that on $K_S^{0}$. In the strange
baryon comparisons, the slope change due to string tension variation shows a
clear hierarchy depending on the strangeness number. The difference between the
slope of minimum string tension case ($\kappa_0$) and maximum string tension
case ($\kappa_{eff}^{s+m}$) in $\Lambda$ production is the smallest, while the
$\Omega$ slope changes most dramatically from $\kappa_0$ case to
$\kappa_{eff}^{s+m}$ case. This strangeness dependent slope change is a direct
outcome as our implementation involves the variation of the effective string
tension which determines the relative production of strange sector in $q\bar{q}$
pair, diquark anti-diquark pair and spin 1 diquark anti-diquark pair at string
breakups.

Aside from examining the integrated yield of particles as a function of event
multiplicity, it is of more interest to understand the relative production of
strange particles. Figure~\ref{fig:BOverM_mult} shows the baryon to meson
production ratio in strange ($\Lambda/K_{S}^{0}$) and non-strange ($p/\pi$) sector.
In the constant string tension scenario represented by the solid line, we
observe no dependence of baryon to meson ratio on event multiplicity. The
inclusion of effective string tension results in a mild increase on both
$\Lambda/K_{S}^{0}$ and $p/\pi$ ratios. This is consistent with observations
in the experimental data. If we set the $\beta$ parameter to higher value,
the variational string tension assumptions may lead to a stronger increasing
trend with event multiplicity.

The strange particle relative to $\pi$ production is shown in
Fig.~\ref{fig:strangeOverPion_mult} for $K_S^{0}$, $\Lambda$, $\Xi$ and
$\Omega$. The rising trend on event multiplicity of the strange particle to
$\pi$ ratio can be explained by the inclusion of string tension variational
scheme in the simulation, while a constant string tension assumption gives a
flat ratio over the whole event multiplicity range. It is also observed that the
gluon wrinkled string structure modification and the multiple string interaction
coordinated method implemented in $\kappa_{eff}^{s}$ and $\kappa_{eff}^{m}$
predict different event multiplicity dependence of the strange to pion ratio.
The relative strange production in $\kappa_{eff}^m$ curve grows monotonously
with the event multiplicity, while the $\kappa_{eff}^s$ curve increases
dramatically in low multiplicity events and then saturates if the mid-rapidity
charged density becomes larger than 4. This observation is consistent with our
knowledge to the multiplicity dependent string tension change in two scenarios
as described in Fig.~\ref{fig:kappa_variation}. The $\kappa_{eff}^{s+m}$ curve
shown with the combined effective string tension change is thus divided into two
domains. The rapid increase in low multiplicity region is dominated by the
impact of gluon wrinkle effect as in $\kappa_{eff}^{s}$. The inter-string
interaction takes over and leads to a mild increase of the strange to pion
ratio in high multiplicity events. Taking account of the strangeness number of
the particles analyzed in this comparison, one can also find the relative
production of multi-strange particles are more sensitive to the event
multiplicity variation than $K_{S}^{0}$ and $\Lambda$. Additionally, one may
also find the multiple string interaction caused effective string tension
modification becomes very small in low multiplicity events. Thus, strange to
pion ratios at $<dN_{ch}/d\eta> \sim 2$ are grouped into two categories,
``$\kappa_{eff}^{m} \approx \kappa_0$'' and ``$\kappa_{eff}^{s+m} \approx
\kappa_{eff}^{s}$''. The single string structure relevant effect comes from the
gluon radiations inside a string object, thus the impact is going to be stronger
with the growth of collision energy~\cite{Long:2011tk}. It is therefore natural
to see the modifications introduced by $\kappa_{eff}^{s}$ or
$\kappa_{eff}^{s+m}$ relative to $\kappa_{0}$ scenario
are still visible even in the low multiplicity events at the LHC energy scale.
This multiplicity dependent strangeness enhancement feature can be well
understood in the
Lund string fragmentation framework by taking the single string structure variation
and multiple string interaction mechanisms into consideration.


\section{Discussions and Conclusions}
\label{sec:summary}
The multiplicity dependent strangeness enhancement observed in pp or p-Pb
collisions is an intriguing finding which triggers a lot of theoretical interest
and continuing experimental investigations for a wider range of energy
dependence. We provide a systematic study on the strange particle productions in
high energy pp collisions based on several variational string tension in
the string fragmentation framework. It turns out
that in the $\kappa_{eff}^{m}$ case, the strange particle yields are generally
in agreement with the experimental data. However,
the $p/\pi$ ratio is still overestimated with the current parameter setup for
all different string tension cases. The multiplicity dependent strangeness
enhancement can be naturally included in the string tension change mechanism.
The multiple string interaction caused effective string tension change can offer
a reasonable starting point for the observed increasing trend of strange flavor
composition, suggesting the multi-string interactions are of great interest to
explain the multiplicity dependence.

In addition, the effective string tension in Eq.~\ref{eqn:global_kappa},
$\kappa_{eff}^{m}$, can also be extended to other collision types by revising
the relevant term to include the nuclear size dependence explicitly. We will
leave the comparisons of the extensions to p-Pb and Pb-Pb to the next work. It
must be noticed that only the flavor composition effect which can be directly handled by
the tunneling mechanism is considered in the current framework. Other
collective-like effects such as the near side ridge in two particle correlations
or the radial flow observed in the transverse momentum distribution of the high
multiplicity pp collisions still require the inclusion of other mechanisms to
deal with the kinematic effect of string dynamics in the high density
environment.

In the end, this study provides a systematic comparison between theoretical and
experimental results of the flavor composition relevant collective behavior in
pp collisions based on the string fragmentation scheme. It is a starting point
for further exclusive studies combining different mechanisms in a more detailed
modeling with microscopic tracing of the parton and hadron evolution history.

\begin{acknowledgments}
This work was supported by the Fundamental Research Funds for the Central
Universities, China University of Geosciences (Wuhan) No.CUG180615, the National
Natural Science Foundation of China (11775094, 11475149) , the Innovative Research Funds
for the Central Universities (2005170503) and the National Key Research and
Development Program of China (2016YFE0100900).
\end{acknowledgments}

\normalem
\bibliography{reference}

\end{document}